\documentclass[aps,prl,twocolumn,showpacs,preprintnumbers,amsmath,amssymb]{revtex4}

\usepackage{graphicx}
\usepackage{dcolumn}
\usepackage{bm}

\input epsf

\usepackage{amsmath}
\usepackage{amsfonts}
\usepackage{amssymb}

\begin{document}

\preprint{APS/123-QED}

\title{Phototactic Clustering of Swimming Micro-organisms in a Turbulent Velocity Field}
\author{Colin Torney}
\author{Zolt\'{a}n Neufeld}
%
\affiliation{School of Mathematical Sciences and Complex \& Adaptive Systems Laboratory\\University College Dublin, \\Belfield, Dublin 4, Ireland}%

\date{\today}

\begin{abstract}
We study the distribution of swimming micro-organisms advected by a model turbulent flow and attracted towards a localised light source through phototaxis. It is shown that particles aggregate along a dynamical attractor with fractal measure whose dimension depends on the strength of the phototaxis. Using an effective diffusion approximation for the flow we derive an analytic expression for the phototactic gain (increase in light exposure over the aggregate) and by extension an accurate prediction for the fractal dimension based on the properties of the advection dynamics and the statistics of the attracting field. This shows that the fractal characteristics of the aggregate are determined by the non-dimensional ratio of the kinetic energy of swimming to that of the turbulent flow. \par
\end{abstract}

\pacs{Valid PACS appear here}
\maketitle

The inhomogeneous distribution of chaotically advected particles has received much recent attention in a variety of contexts. It has been shown in the case of passive particles in compressible surface flows \cite{ott_sommerer,boffetta,boffetta2} and inertial particles in incompressible flow \cite{bec,bec2,Duncan} that clusters form along a dynamical attractor of fractal measure. This intermittent distribution has important consequences for physical process such as rain formation \cite{falkovich} and chemical or biological reaction processes \cite{karolyi2,karolyi}. In the context of biological processes microscale patchiness in plankton populations has been observed in several studies \cite{tsuda,seuront,mitchell,mitchell2} and is seen to play an important role in plankton population dynamics and therefore oceanic ecology.

\par

In this letter we consider swimming particles advected by an incompressible flow with an effective compressibility induced by particle motility in the presence of an attracting light source. Particles are able to detect the gradient of the illumination field and move in the direction of increasing light intensity with a swimming velocity assumed to be proportional to the detected gradient. Motion is therefore governed by the following equation
\begin{equation}
\mathbf{\dot{r}} =  \mathbf{{v}}_{f}({\bf r},t)+\chi\nabla \Phi({\bf r})
\label{eqn_1} 
\end{equation} 
where the parameter $\chi$ is the phototactic coefficient which defines the strength of the particles reaction to the gradient of the illumination field $\Phi({\bf r})$. For our simulations we assume a Gaussian light distribution although results are not dependent on the exact functional form of $\Phi({\bf r})$. \par
As a carrier flow we use a two-dimensional synthetic turbulence model which approximates a turbulent flow \cite{careta,stuart}. It is Gaussian, isotropic and statistically homogeneous with a prescribed energy spectrum of the form
\begin{equation}
E(k) \varpropto k^3 \exp\left[ -\frac{k^2}{k_0^2}\right] 
\label{kraich}
\end{equation} 
originally proposed by Kraichnan \cite{kraichnan}. The flow is generated by using an Ornstein-Uhlenbeck process defined as the solution to the stochastic partial differential equation
\begin{equation}
\frac{\partial\psi}{\partial t}=\nu \nabla^2 \psi + \sqrt{\xi}\frac{\partial{W}}{\partial t}
\end{equation} 
where $\psi$ is the stream function of the velocity field, $\nu$ is a viscosity parameter that sets the time correlation of the flow, $\xi$ represents the magnitude of the stochastic forcing and $W$ is a Wiener process. Following \cite{stuart} we introduce the lengthscale L and timescale T such that $T=L\sqrt{v/\xi}$ and rescale so that the flow is periodic on a square domain with unit length. The lengthscale of the highest energy modes is defined by $2\pi/k_0$ and we set this value to be two thirds of the box size. The spectrum is normalised which leads to a mean squared velocity of the flow of 0.5. For the illumination field we use a Gaussian function with an amplitude of unity and a width of a quarter of the box length, $\Phi({\bf r})=\exp{\left[ -8\left(x^2+y^2\right)/L^2\right]}$.
\par
\begin{figure}
\includegraphics[scale=0.56,clip]{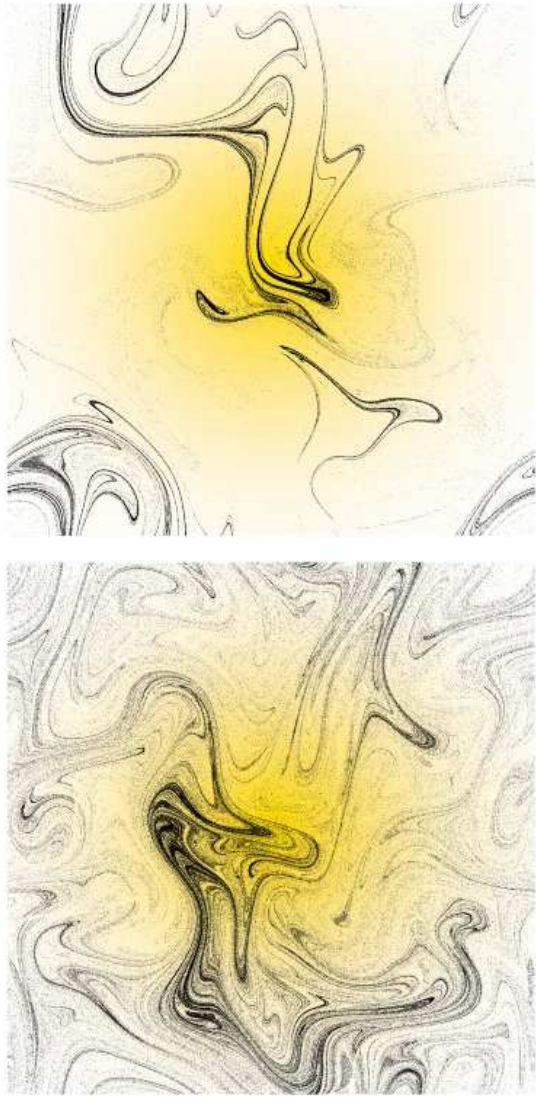}
\caption{Distribution of 500,000 phototactic particles after relaxation of transients. Above: $\chi=0.15$. Below: $\chi=0.05$. Color version: Light distribution is shown in yellow}
\label{par_dist}
\end{figure}

In an ergodic incompressible flow with no transport barriers an ensemble of passively advected particles are quickly dispersed producing a spatially uniform density distribution. Conversely phototactic particles swimming in the direction of a fixed illumination gradient in the absence of advection would concentrate in regions of maximum light intensity. When combined these two effects result in a statistically stationary steady state particle distribution which is highly non-uniform in space. Particle distributions after the relaxation of transients are shown in this regime in Fig.~\ref{par_dist}. Particles clearly do not converge to a fixed point as would be the case for strong phototactic attraction but instead aggregate along a dynamically evolving fractal attractor and continue to visit all regions of the physical space.\par

\begin{figure}
\includegraphics[angle=-90,scale=0.44,clip]{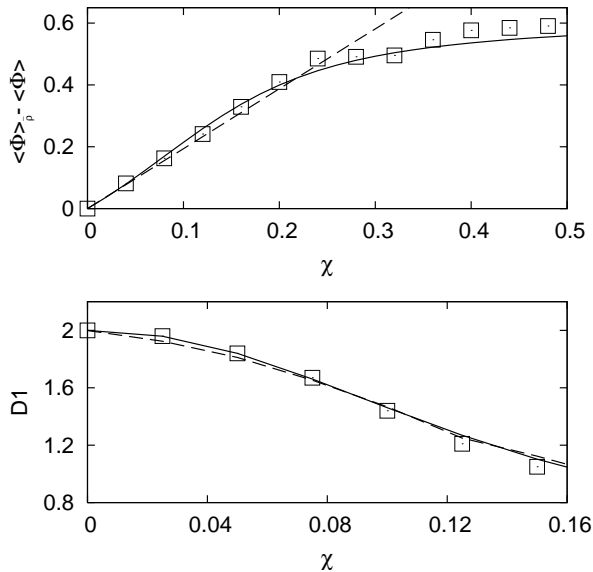}
\caption{Above: Gain in light exposure achieved by phototaxis. Squares - numerical results, solid line - theoretical approximation, dashed line - first order theoretical approximation. Below: Information dimension. Squares - values calculated from distribution by Eqn.~\ref{eqn_D1}, dashed line - results from numerically calculated Lyapunov exponents, solid line - theoretical model.}
\label{fig2}
\end{figure}

Despite this, the average illumination received by the ensemble of particles (or equivalently along the trajectory of a single particle) increases with $\chi$ and is higher than the spatial average illumination. In Fig.~\ref{fig2} we show the time-averaged increase in light exposure of an ensemble of phototactic particles as compared to passively advected non-motile particles as a function of the phototactic coefficient.\par

As $\chi$ is increased the attractor becomes more singular with most particles concentrated in thin filaments separated by empty regions. To quantify the characteristics of the attractor we numerically calculate the information dimension defined as \cite{ott_book}
\begin{equation}
D_1 =\lim_{\epsilon\to 0} \frac{ \sum\limits_{i=1}^{\tilde{N}(\epsilon) }\mu_i \ln{\mu_i}  } {\ln{\epsilon}}
\label{eqn_D1} 
\end{equation} 
where the domain is covered by a rectangular grid of box size $\epsilon$ and $\mu_i$ is the fraction of particles in a given box $i$. Results for various values of $\chi$ are shown in Fig.~\ref{fig2}.

\par


In order to investigate the relationship between the phototactic coefficient $\chi$ and both the average illumination gain of the aggregate and the fractal dimension of the attractor we approximate the turbulent advection as a purely diffusive process. 
\par
\begin{figure}
\includegraphics[angle=-90,scale=0.33,clip]{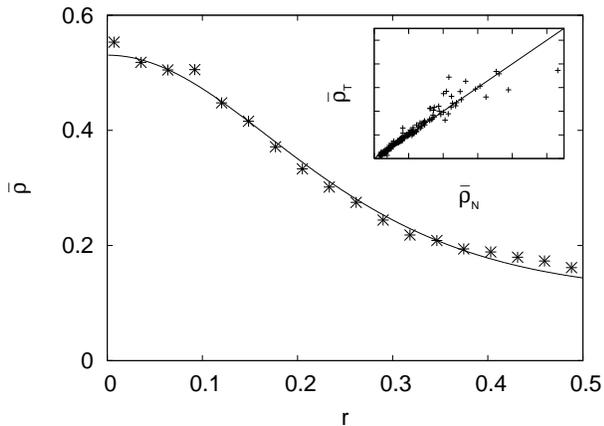}
\caption{Radial plot of probability density function (solid line) and numerical results (points) for average particle distribution, $\chi=0.05$. Inset: Scatter plot of predicted density ($\bar{\rho}_T$) to numerically observed density ($\bar{\rho}_N$) for range of $\chi$ from 0.025 to 0.15. Points lying on the diagonal represent exact match between effective diffusion model and numerics.}
\label{fig3}
\end{figure}

Although the instantaneous dynamical attractor changes in an irregular fashion following the turbulent flow, by averaging over time we can obtain a picture of the denser regions of the physical space which will then allow us to quantify various statistical properties of the system. To this end we define $\bar{\rho}$ as the time averaged density of particles at a given point in space and the notation $\langle .. \rangle_{\bar{\rho}}$ to mean an average quantity defined over $\bar{\rho}$. With these definitions and an effective diffusion approximation for the flow we model the system by the equation
\begin{equation}
\frac{\partial \bar{\rho} }{\partial t}   = D_{f} \nabla^2\bar{\rho} + \nabla \cdot (\bar{\rho} \chi\nabla \Phi)
\label{eq_adv}
\end{equation} 
where $D_f$ is the effective diffusivity of the flow defined by the asymptotic dispersion rate of advected passive particles in an unbounded domain as $\langle d^2 \rangle \simeq 4D_ft$. The first term on the r.h.s. of Eqn.~\ref{eq_adv} represents the diffusive effect of the turbulent flow which smooths out any inhomogeneities in the distribution and if dominant would lead to a uniform particle density. The second term counteracts this effect by collecting particles in high light intensity regions leading to a non-uniform distribution. Eqn.~\ref{eq_adv} has a steady state solution of the form
\begin{equation}
\bar{\rho}(\mathbf{r})= \frac{1}{Z}\exp\left[ {\frac{\chi}{D_f} \Phi(\mathbf{r})}\right] 
\label{eq_sol}
\end{equation} 
which is analogous to a Boltzmann distribution of non-interacting particles in an external field, where $Z$ is a normalisation factor akin to a partition function which allows us to use $\bar{\rho}$ as a probability density
\begin{equation}
Z= \int^{V}  \exp\left[ {\frac{\chi}{D_f} \Phi(\mathbf{r})}\right]dV.
\label{eq_Z}
\end{equation}
To compare this result with the numerical experiments we first note that as a result of the radial symmetry of the illumination field the particle distribution is also radially symmetric. This allows us to fully describe the two dimensional distribution by a one dimensional radial profile. In Fig.~\ref{fig3} the result predicted from Eqns.~\ref{eq_sol} and ~\ref{eq_Z} is compared to a time averaged, normalised histogram plot of radial location of particles for $\chi=0.05$. We also include in the inset of Fig.~\ref{fig3} a scatter plot of numerical results against expected values for a range of $\chi$ which illustrates the excellent agreement of numerics with the effective diffusion approximation.
\par

Using the distribution given above we can evaluate the average light intensity experienced by particles with a given phototactic response parameter $\chi$, as 
\begin{equation}
\left\langle \Phi \right\rangle_{\bar{\rho}}=\frac{1}{Z}\int^{V}\exp{\left[ \frac{\chi}{D_f}\Phi(\mathbf{r}) \right]} \Phi(\mathbf{r})dV 
\label{eq_av_c}
\end{equation}
which can be rewritten as 
\begin{equation}
\left\langle \Phi \right\rangle_{\bar{\rho}}  = D_f {\frac{\partial \log{Z}}{\partial \chi}}.
\end{equation}
Then as $Z$ is an expectation value over the unit area domain the logarithmic term can be treated as a cumulant generating function of the light distribution $\Phi$. This leads to
\begin{equation}
\left\langle \Phi \right\rangle_{\bar{\rho}} = \sum_{n=0}^{\infty} \kappa_{n+1} \frac{1}{n!} \left( \frac{\chi}{D_f} \right)^n  
\label{eq_av_c3}
\end{equation}
where $\kappa_n$ are the cumulants of the illumination field. For weakly phototactic particles, i.e. $\chi/D_f\ll1$, the higher order terms can be neglected leaving
\begin{equation}
\left\langle \Phi \right\rangle_{\bar{\rho}} \simeq \langle \Phi \rangle + \frac{\chi}{D_f}\left( \left\langle \Phi^2 \right\rangle -  {\left\langle \Phi \right \rangle}^{2}\right).
\end{equation}
This explains the linear relationship seen in Fig.~\ref{fig2} for small values of $\chi$ and shows that the gain in illumination exposure due to phototaxis is proportional to the variance of the illumination field. The truncated series is plotted in Fig.~\ref{fig2} alongside results containing higher order terms required for convergence.
\par

Interestingly, the time-averaged distribution defined by Eqn.~\ref{eq_sol} can also be used to calculate the dimension of the dynamical attractor. The Kaplan-Yorke conjecture~\cite{kaplan} relates the information dimension of an attractor to the average stretching and contraction rates in the phase space by the formula (for two dimensional phase space)
\begin{equation}
{D}_1 = 1 + \frac{{\lambda}_1}{\vert{{\lambda}_2}\vert}
\label{eq_ky}
\end{equation}
where $\lambda_1$, $\lambda_2$ are the Lyapunov exponents of the system. The sum of the Lyapunov exponents defines a dissipation rate $\alpha$, i.e. the average rate at which volumes in the phase space contract. In a conservative system as is the case of particles passively advected by an incompressible flow $\alpha$ is zero and therefore in the two dimensional model $\lambda_1=-\lambda_2$ and the information dimension is two. However if phototaxis is present particles no longer follow the streamlines of the flow and there is a net contraction in the system which reduces the dimension of the attractor. 
\par
\begin{figure}
\includegraphics[angle=-90,scale=0.33,clip]{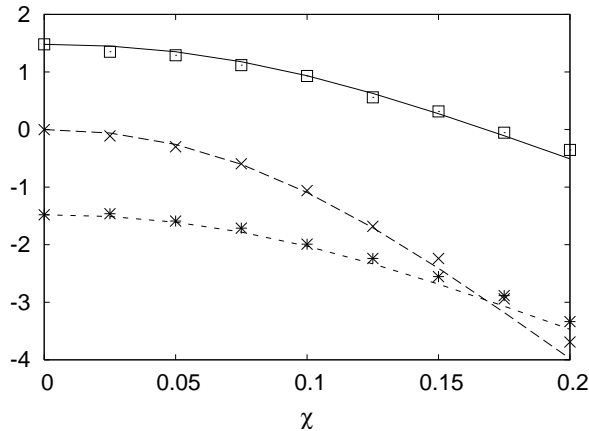}
\caption{Lyapunov exponents and dissipation rate. Symbols represent numerical values found by the Benettin method, lines values calculated from the theoretical approximation defined by Eqns.~\ref{eq_av_nabc} and~\ref{eq_exps} : Squares/solid line - $\lambda_1$, crosses/ heavy dashed - $\alpha$, asterisks/fine dashed - $\lambda_2$.}
\label{fig4}
\end{figure}

The Lyapunov exponents can be determined numerically by a standard method proposed by Benettin~\cite{benettin} and the values found in this way show the dependence of $\lambda_1$, $\lambda_2$ and $\alpha$ on the phototactic coefficient (Fig.~\ref{fig4}). The information dimension based on these quantities can now be found by using the Kaplan-Yorke conjecture and this agrees with the values calculated directly from the particle distributions as shown in Fig.~\ref{fig2}.
\par

Returning to the theoretical model the dissipation rate $\alpha$ can be obtained from the divergence of the velocity field of the phototactic particles. Due to the incompressibility of the carrier flow this gives
\begin{equation}
\alpha=\left\langle \chi \nabla^2 \Phi \right\rangle_{\bar{\rho}}=\int^{V}\bar{\rho} \chi\nabla^2\Phi(\mathbf{r}) dV 
\label{eq_av_nabc}
\end{equation}
where by ergodicity averaging over a single trajectory over long times is taken to be equivalent to the weighted average over the distribution $\bar{\rho}$ previously defined in our continuum description.\par

From the isotropic properties of the flow and zero correlation between fluid dynamics and illumination field we expect the contraction of phase space elements due to phototaxis to have an equal effect on both Lyapunov exponents. This is consistent with numerical results as shown in Fig.~\ref{fig4}. Defining $\lambda_0$ as the positive exponent of the carrier flow (i.e. in the case of passive particles $\lambda_1=\lambda_0$, $\lambda_2=-\lambda_0$) the exponents for the phototactic particles are then given by
\begin{eqnarray}
\lambda_1=\lambda_0 + \frac{1}{2}\left\langle \chi\nabla^2 \Phi \right\rangle_{\bar{\rho}} &&
\lambda_2=-\lambda_0 + \frac{1}{2}\left\langle \chi\nabla^2 \Phi \right\rangle_{\bar{\rho}} .
\label{eq_exps}
\end{eqnarray}
For the Gaussian light distribution the integral in Eqn.~\ref{eq_av_nabc} can be evaluated numerically and values for $\lambda_1$, $\lambda_2$ and $\alpha$ are plotted in Fig.~\ref{fig4} alongside those obtained from the Benettin method. Combining Eqns.~\ref{eq_ky},~\ref{eq_av_nabc} and~\ref{eq_exps} gives an expression for the fractal dimension of the distribution in terms of $\Phi$, $\chi$ and the properties of the carrier flow field $\lambda_0$ and $D_f$. 
\begin{equation}
D_1=\left[\frac{1}{2}-\frac{1}{4{\lambda_0}Z}\int^{V}\exp{\left[ \frac{\chi}{D_f}\Phi\right]} \chi\nabla^2\Phi dV\right] ^{-1} .
\label{total_D1}
\end{equation}
To illustrate their agreement with the numerical results solutions to this equation are included in Fig.~\ref{fig2}.
\par
An alternative formulation can be obtained from Eqn.~\ref{eq_av_nabc} by the use of integration by parts 
\begin{equation}
\alpha=     \oint^S \bar{\rho}\chi \nabla\Phi dS - \int^{V} \frac{\bar{\rho}}{D_f} \left( \chi\nabla\Phi\right) ^2 dV
\label{alpha_by_parts}
\end{equation}
where the boundary term vanishes for most boundary conditions (i.e. smooth periodic or no-flux illumination field)~\cite{note}. The second integral demonstrates that $\alpha$ is always negative and is proportional to the mean square swimming velocity of the phototactic particles. Thus the information dimension can be written as

\begin{equation}
D_1=2\left[1+\frac{ \langle{V_s^2}\rangle}{2{\lambda_0}{D_f}} \right] ^{-1} .
\label{vs_D1}
\end{equation}
In a fully chaotic flow the effective diffusivity can be estimated as $D_f\sim UL$ and the Lyapunov exponent is an inverse characteristic timescale of the flow $\lambda_0\sim U/L$. Using these relations Eqn.~\ref{vs_D1} shows that the information dimension is a function of the non-dimensional ratio of the average kinetic energy of swimming measured relative to that of the carrier flow.
\par
We have therefore seen how the effects of phototaxis on the spatial distribution and increase in light exposure of an ensemble of micro-organisms can be explained by an analysis of the attracting field and the characteristics of the carrier flow. Specifically we have related the dimensionality of the distribution to the measurable kinetic energy of the swimming organisms. 
\par
The combined effect of phototaxis and turbulent flow leads to highly non-uniform distribution of micro-organisms and this provides a plausible mechanism for the generation of microscale patchiness observed in experiments. Our analytical results for the key characteristics of systems of phototactic particles may be used to make testable quantitative predictions or to extract information from measurements (e.g. based on the relationship between swimming kinetic energy and dimension).
\par


\begin{thebibliography}{20}
\bibitem[{not()}]{note}
\bibinfo{note}{Note: This condition is not exactly satisfied in our numerical
  simulations as the Gaussian illumination field has a small discontinuity in
  the gradient at the boundaries.}
\end{thebibliography}

\end{document}